\documentstyle[preprint,aps]{revtex}
\begin{document}
\def\hw {\hbar \omega}
\def\ie{{\em i.e., }}
\def\rf#1{{(\ref{#1})}}
\def\sss{\scriptscriptstyle}
\def\etal{{\it et al.}}
\def\cf{{\it cf }}
\def\mev{\; {\rm MeV} }
\draft
\title{Suppression of core polarization in halo nuclei }

\author{T.T.S. Kuo${}^1$, F. Krmpoti\'c${}^{2}$, and Y. Tzeng${}^3$}
\address{${}^1$Department of Physics, SUNY-Stony Brook\\
Stony Brook, New York 11794 USA\\
${}^2$
Departamento de F\'\i sica, Facultad de Ciencias Exactas\\
Universidad Nacional de La Plata, C. C. 67, 1900 La Plata, Argentina\\
${}^3$ Institute of Physics, Academia Sinica \\
Nankang, Taipei, Taiwan}
\date{\today}
\maketitle

\begin{abstract}
We present a microscopic study of halo nuclei, starting from the Paris and
Bonn potentials and employing a two-frequency shell model approach. It is
found that the core-polarization effect is dramatically suppressed in such
nuclei. Consequently the effective interaction for halo nucleons is almost
entirely given by the bare $G$-matrix alone, which presently  can be evaluated
with a high degree of accuracy. The experimental pairing energies between
the two halo neutrons in $^6\!He$ and $^{11}\!Li$
nuclei are satisfactorily reproduced by our calculation.
It is suggested that the fundamental nucleon-nucleon interaction can be
probed in a clearer and more direct way in halo nuclei than
in ordinary nuclei.
\end{abstract}

\pacs{PACS numbers: 21.30.+y,21.60.Cs, 21.90.+f}

\newpage
\narrowtext

Radioactive-beam nuclear physics has been progressing rapidly, and there
is much current interest in studying halo nuclei \cite{ber95,al96}.
There were four articles about
halo nuclei
in a recent issue of Physical Review C:
Nazarewicz \etal  \cite{naz96} dealt with the halo nuclei
around the nucleus $^{48}\!Ni$, Hamamoto \etal \cite{ham96} carried
out a systematic investigation of the single particle and collective
degrees of freedom in the drip-line nuclei, and an
experimental study of heavy halo nuclei around $N=82$ was also presented
\cite{pag96}.
It is remarkable that nuclei as exotic as  $^{48}\!Ni$, \ie the mirror
image of the double closed-shell nucleus $^{48}\!Ca$, are now being
studied!

The halo nuclei (or drip-line nuclei) may well play a central role in
our understanding of the nuclear binding. Their typical structure is that
of a tightly bound inner core with a few outer nucleons that are loosely
attached to the core. Although these exotic nuclei are bound,
their binary subsystems are not. For instance, the halo nucleus $^6\!He$
($^{11}\!Li$) is presumably made of a $^4\!He$ ($^{9}\!Li$) core surrounded
by a two-neutron halo. As a whole $^6\!He$ ($^{11}\!Li$) is bound, but its
binary subsystems, \ie $^5\!He$ ($^{10}\!Li$) and the {\it di-neutron} are
unbound.
The pairing force between the valence nucleons is thus essential
for the stability of the halo nuclei,
and it is important to calculate it as accurately as we can.

So far, the halo nuclei have been calculated using empirical effective
interactions, tuned to stable nuclei. The inherent density dependence of
Skyrme-type forces \cite{naz96,ham96} provides a
reasonable means of extrapolating to the lower density regimes
characteristic of nuclei far from stability. Yet, quite recently Kuo
\etal \cite{kuo96} have suggested to study the drip-line nuclei
from the first principles, \ie from the elementary nucleon-nucleon
(NN) force, such as the Paris \cite{lac80} and Bonn \cite{mac89} interactions.
Halo nucleons are separated rather far from the other nucleons
in the "core nucleus", and the interaction among them should be derivable
from the free NN interaction with small medium corrections.

Starting from a free
NN interaction,  a model-space effective interaction
($V_{\sss eff}$) among the nucleons in the nuclear medium can be derived,
using a $G$-matrix folded-diagram approach \cite{kuo90,kuo95}.
The major difficulty in such a microscopic effective interaction theory
has been the treatment of the core polarization effect (CPE), in particular
the higher-order core polarization diagrams.
In ordinary nuclei, the valence nucleons are  close to the nuclear core,
an example being the two $sd$-shell neutrons of $^{18}\!O$ residing
adjacently to the $^{16}\!O$ core. Consequently, there is a strong
valence-core coupling and therefore a large CPE.
It is a formidable task  to figure out in such a situation which $G$-matrix
diagrams should be embodied in the $\hat{Q}$-box. The two leading terms
are the well-known first-order $G$-matrix diagram and the second-order
core polarization diagram $G_{3p1h}$ \cite{shu83}. Hjorth-Jensen \etal
\cite{jorth} have investigated the third-order $\hat{Q}$-box diagrams
for the $sd$ shell. They concluded that, after folding, the net effect on
$V_{\sss eff}$ was a change of about $10-15\%$, as compared with the case
when only the first- and second- order ones were considered.
Higher and higher-order core polarization diagrams rapidly become
prohibitively more difficult to deal with. Thus in practice one can only
include some low-order diagrams for the calculation of the $\hat{Q}$-box.
Besides, when there
are disagreements between  theory and experiment, one is not sure if they
are due to the NN interaction or to the approximation adopted  in
solving the many-body problems (such as the neglecting of the  higher-order
core polarization diagrams).
The environment in the halo nucleons is different and may be more promising,
because they are located quite far away from the core.
A schematic comparison between normal and halo nuclei is given in Fig. 1.
A relatively weak CPE is expected for halo nucleons,
and therefore an
$V_{\sss eff}$
predominately governed by the free NN interaction.
That is, $V_{\sss eff}$ should be in essence the bare $G$-matrix,
which presently can be calculated to a high degree of accuracy.
Hence the halo nuclei, besides  having excitingly interesting and exotic
properties, may furnish  as well a much better testing ground for the
fundamental NN interactions, than ordinary nuclei.

Motivated by  the above scenario, we present in
this letter a microscopic derivation of the $V_{\sss eff}$
for halo nucleons in $^6\!He$ and $^{11}\!Li$, starting from the Paris and
Bonn NN potentials and using a $G$-matrix folded-diagram approach
\cite{kuo90,kuo95}. The main steps in such a derivation are:

i) {\it Choice of the model space $P$}.
An important criterion for selecting the model space $P$ is that its
overlap, with the physical states under consideration, should be as large
as possible.
For instance, the $^4\!He$, i.e., the $^6\!He$ core, should remain
essentially as an ordinary $\alpha$-particle, with little perturbation
from the distant halo nucleons.
For the $P$ space
we shall use a closed $(0s_{1/2})^4$ core ($\alpha$-particle) with the
valence (halo) nucleons confined in the $0p$ shell. Yet the halo nucleons
have a much larger r.m.s. radius than the core, and therefore an oscillator
constant $\hbar\omega$ considerable smaller than that given by the
empirical formula $\hbar\omega=45A^{-1/3}-25A^{-2/3}$ MeV (valid for
ordinary nuclei). It would not be then feasible to reproduce both radii,
using shell model wave functions with a common $\hbar\omega$.
One may get past this difficulty by including several major shells in the
one-frequency shell model (OFSM) calculation. But this would be very tedious.
A convenient and physically appealing solution to this problem is to employ a
two-frequency shell model (TFSM) for the description of halo nuclei, as
suggested in Ref. \cite{kuo96}.
Within the TFSM one uses oscillator wave functions with $\hbar\omega_{in}$
and $\hbar\omega_{out}$ for the core (inner) and the halo (outer) orbits,
respectively. The notations $b_{in}$ and $b_{out}$ also will be used
from now on, with $b^2\equiv \hbar/m\omega$. In the present work
$b_{in}$ is fixed at $1.45$ fm,  while $b_{out}$ is treated as a variation
parameter (or generator coordinate). To assure the orthonormality, we have
actually used $b_{in}$ for all the $\ell=0$ waves ($0s_{1/2},1s_{1/2},\cdots$)
and $b_{out}$ for waves with other $\ell$ values.

ii) {\it Evaluation of the model-space $G$-matrix.}
For ordinary nuclei, the $G$-matrix can be calculated rather accurately with
the method developed in Refs. \cite{tsa72,kre76}. We extend below this method
to the halo nucleons in the context of the TFSM.
For a general  model-space $P$, we define the corresponding Brueckner
$G$-matrix by the integral equation \cite{kre76,mut92}
$$
G(\omega)=V+V Q_2\frac{1} {\omega-Q_2TQ_2}Q_2G(\omega),
$$
where $\omega$ is an energy variable, $Q_2$ is a two-body Pauli exclusion
operator, and $T$ is the two-nucleon
kinetic energy. Note that our $G$-matrix has orthogonalized plane-wave
intermediate states. The exact solution of this $G$-matrix is
$G=G_F+\Delta G$ \cite{tsa72,kre76}, where $G_F$ is the {\it free} $G$-matrix,
and $\Delta G$ is the Pauli correction term
$$
\Delta G(\omega)=-G_F(\omega)\frac{1}{e}P_2\frac{1}
{P_2[\frac{1}{e}+\frac{1}{e}G_F(\omega)\frac{1}{e}]P_2}
P_2\frac{1}{e}G_F(\omega),
$$
with $e\equiv \omega-T$. The projection operator $P_2$, defined as
$(1-Q_2)$, will be discussed later.
The basic ingredient for calculating the above $G$-matrix is the matrix elements
of $G_F$ within the $P_2$ space. This space contains all
the two-particle states that must be excluded from the intermediate states
in $G$-matrix calculations. For ordinary nuclei, where the OFSM is used, the
states excluded by the Pauli operator and those contained within the model
space have a common length parameter $b$.
For halo nuclei, where we use the TFSM, the situation is more complicated,
as the wave functions
for the excluded states and
those within the model space have in general
different length parameters
$b_{in}$ and $b_{out}$.
Hence to calculate $\Delta G$, we need the matrix elements of $G_F$
in a $b_{in}-b_{out}$ mixed representation. This poses a technical
difficulty because the transformations, from the c.m. coordinates to
the laboratory coordinates for  two-particle states with different
oscillator lengths, are not as easy to perform as for one common oscillator
length. We have adopted an expansion procedure to surmount this difficulty.
Namely, we expand the oscillator wave functions with $b_{in}$ in terms of
those with $b_{out}$, or {\it vice versa}. When $b_{in}$ and $b_{out}$ are
not too different from each other, this procedure is relatively effortless
to carry out.
Usually a high accuracy can be attained by including about $8$ terms in the
expansion. Still, the calculation of the two-frequency $G$-matrix is
significantly more complicated than the ordinary one-frequency one.
Another difficulty, in deriving the $G$-matrix for halo nuclei, is
the treatment of its Pauli exclusion operator. As the halo nucleons are
rather far from the core nucleons, the effect of Pauli
blocking is expected to be small.
But, to get a reliable result for a small effect,
a very accurate procedure has to be employed.
We write the projection operator $Q_2$ as
$$
Q_2=\sum_{all~ab} Q(ab)\vert ab\rangle \langle  ab \vert,
$$
where $Q(ab)=0$, if $b\leq n_1,~a\leq n_3$, or $b\leq n_2,~a\leq n_2$,
or $b\leq n_3,~a\leq n_1$, and $Q(ab)=1$ otherwise.
The boundary of $Q(ab)$ is specified by the orbital numbers ($n_1,n_2,n_3$).
We denote the shell model orbits by numerals, starting from the bottom of
the oscillator well: $1$ for orbit $0s_{1/2}$, $2$ for $0p_{3/2},\cdots, 7$
for  $0f_{7/2}$, and so on. $n_1$ and $n_2$ stand for the highest orbits
of the closed core (Fermi sea) and of the chosen model space, respectively.
For example, we consider $^4\!He$ as a closed core and all $6$ orbits in
the $sp$ and $sd$ shells are included in the model space. Then $n_1=1$ and
$n_2=6$.
As for the $G$-matrix intermediate states we consider
only particle states (\ie states above the Fermi sea),
$n_3$ in principle should be $\infty$ \cite{kre76}.
Still, in practice
this is not
feasible, and $n_3$ has to be determined by an empirical
procedure. Namely, we perform calculations with increasing values for $n_3$
until  numerical results become stable.
In Table 1, we display some representative results of our two-frequency
$G$-matrix for the $\{0s0p\}$ model space, with $b_{in}=1.45$ fm and
$b_{out}=2.0$ fm. The only approximation here is the finite $n_3$ truncation.
A satisfactory $n_3$ convergence is attained for $n_3=21$, and this
value is used here. It is worth noting that,
although the halo nucleons are widely separated from the closed core,
the Pauli correction term $\Delta G~(=G-G_F)$ is still quite significant.

iii) {\it Calculation of the irreducible diagrams for the vertex function
$\hat Q$-boxes}.
Once derived the  $G$-matrix, we can calculate the irreducible vertex function
$\hat Q$-box. Finally, the model-space energy-independent $V_{\sss eff}$ is
evaluated in terms of the $\hat Q$-box folded-diagram series
\cite{kuo90,kuo95}, following closely the procedures of Ref. \cite{shu83}.

Diagonal matrix elements of $G$, $G_{3p1h}$ and $V_{\sss eff}$, for
the states $|(p_{3/2})^2;T=1,J=0\rangle$ and $|(p_{1/2})^2;T=1,J=0\rangle$,
are shown in Fig. 2 as a function of $b_{out}$, for both the Paris and
Bonn-A potentials.
As we increase $b_{out}$, we are augmenting the average
distance between the halo nucleons and the core and so reducing
the coupling between them. For sufficiently large $b_{out}$, the total
CPE must be small and it should be sufficiently accurately given by the
second-order (lowest order)  core polarization diagram alone.
In fact, as $b_{out}$ increases, the core polarization diagrams $G_{3p1h}$
approach rapidly and monotonically to zero and become negligibly
small at $b_{out}\cong 2.25$ fm. In our TFSM approach, we have assumed a
fixed $^4\!He$ or $^{9}\!Li$ core, always described by $b_{in}=1.45$ fm.
Therefore the energy denominator for the diagram $G_{3p1h}$ is fixed by the
corresponding core and does not change with $b_{out}$. This means that
the suppression of $G_{3p1h}$ is entirely due to the weakening of the
core-valence particle interaction.
The behavior of the bare $G$-matrix and $V_{\sss eff}$  shown in
Fig. 2 are also of interest. First, they are quite similar to each other.
Second, while for the $p_{3/2}$ case, they become weaker as $b_{out}$
increases, in the $p_{1/2}$ case they become stronger as $b_{out}$ increases.
Third, at large $b_{out}$ the results given by the Paris and Bonn A
potentials are practically identical. This is because their long-range parts
do not differ much from each other.

To assess to which extent the nuclear model formulated above is reliably
it is necessary to compare our results with experiments. The valence or pairing
interaction energy between the halo nucleons in $^6\!He$ is obtained from the
odd-even mass difference \cite{tul95}
$$E_{p}^{exp}(^6\!He)=-[{\cal B}(^6\!He)+{\cal B}(^4\!He)-2{\cal B}(^5\!He)]\!
=\!-2.77 \mev.$$
To calculate this energy, we need to diagonalize the folded-diagram $V_{eff}$
\cite{shu83} in a  $p_{3/2}$ and $p_{1/2}$ model space ($T=1,J=0$). In this
way we get $E^{th}_{p}=-2.97$ MeV at $b_{out}=2.25$ fm for the Paris potential.
As the CPE is strongly suppressed for such a $b_{out}$ value, this result
almost entirely comes from the bare $G$-matrix.
The ground-state wave function of $^6\!He$ is almost pure $(p_{3/2})^2$,
with very little $(p_{1/2})^2$  admixture. Thus our $E^{th}_{p}$ is also
close to the diagonal $G$-matrix element shown in Fig. 2.
As we have used a $p_{3/2}- p_{1/2}$ model space, our wave function
for $^{11}\!Li$ has only one component (neutron orbits closed). Then
the diagonal $(p_{1/2})^2(T=1,J=0)$ matrix element of $V_{\sss eff}$,
which is quite close to the unfolded value of Fig. 2 at $b_{out}=2.25$ fm,
is directly comparable to the pairing energy for $^{11}\!Li$. From the
masses of $^{11}\!Li$, $^{10}\!Li$ and  $^9\!Li$ \cite{tul95},
we get $E^{exp}_{p}=-1.14$ MeV, while our result is $E^{th}_{p}=-0.81$
MeV at $b_{out}=2.25$ fm for the Paris potential.
This discrepancy could
be pointing out that some physics is still missing in our description of
the valence $^{11}\!Li$ neutrons. It is very likely that they should
not be entirely confined to the $p$ shell, but a larger space, such as
$\{0p_{1/2}0d1s\}$, is probable needed.

In passing, we mention that the above $b_{out}=2.25$ fm  is a
reasonable choice. Recall that we have fixed $b_{in}=1.45$ fm. With these
values of $b_{in}$ and $b_{out}$, and assuming a pure $s^4p^n$ wave function,
we get that $R^{th}(^6\!He)=2.51$ fm, in good agreement with
the empirical value $R^{exp}(^6\!He)=2.57\pm 0.1$ fm \cite{zhu93}; similarly,
$R^{th}(^{11}\!Li)=3.03$ fm while $R^{exp}(^{11}\!Li)=3.1\pm 0.1$ fm
\cite{zhu93}.

We have also calculated the valence interaction energy for $^6\!Li$
using a similar folded-diagram procedure in the $p_{3/2}-p_{1/2}$
space. From the empirical masses
of $^6\!Li$,  $^5\!Li$,  $^5\!He$ and $^4\!He$ \cite{tul95}, we obtain
$E^{exp}_{v}=-6.56$ MeV. (This number was incorrectly given as $-3.55$ MeV
in Ref.\cite{kuo96}.) Our result is $E^{th}_{v}=-6.64$ MeV
for the Paris potential, if we use $b_{out}=1.75$ and $b_{in}=1.45$ fm.
It is of interest to stress that $^6Li$ is {\it not}
a halo nucleus, according  to our calculation, in the sense that there is
no need to employ a  very large $b_{out}$ for its valence nucleons.

In summary, we have derived the effective interaction for the
valence nucleons in halo nuclei, starting from realistic NN interactions.
Our preliminary results are encouraging. We have employed a two-frequency
shell model approach, to give a good spatial description for both the core
nucleons and the halo nucleons. While keeping the inner length parameter
$b_{in}$ fixed, we gauge the spatial extension of the halo nucleons by
varying the outer length parameter $b_{out}$. In this way we have explicitly
proved that the core polarization effect is strongly suppressed at large
$b_{out}$ values, as required by the large  empirical r.m.s. radii of halo
nuclei. Ergo, {\it the effective interaction between the
halo nucleons is predominantly given by the bare $G$-matrix alone}, in accord
with our expectations.
The Pauli blocking effect on the $G$-matrix has been found to be
very important, and it can be calculated quite accurately as we have
demonstrated.
Thus it appears that one can derive the effective interaction
for halo nuclei much more accurately than for ordinary nuclei.
We enthusiastically believe that the halo
nuclei, which have already greatly enhanced our knowledge about nuclei,
may in addition  provide  a more accurate testing ground for the fundamental
NN interaction, than the ordinary nuclei.

\acknowledgments
This work is supported in part by  the USDOE Grant DE-FG02-88ER40388,
by the NSF Grant and by the Fundaci\'on Antorchas (Argentina).
One of us (T.T.S. Kuo) is grateful for the warm hospitality extended to him
while visiting the Universidad Nacional de La Plata.

\begin{table}
\caption{Dependence of the two-frequency $G$-matrix on the choice of $n_3$.
Listed are the matrix element $\langle (0p_{3/2})^2;TJ\vert G(\omega)
\vert (0p_{3/2})^2;TJ \rangle$ (in MeV), calculated for the Paris potential
and three different values of $\omega$ (in MeV),
with $TJ=01$ (upper panel) and with $TJ=10$ (lower panel). We have used
$b_{in}=1.45$  and $b_{out}= 2.0$ fm for the length parameters, and $n_1=1$
and $n_2=6$ for the exclusion operator. The first row in each group (F)
denotes the free $G$-matrix.}
\begin{tabular}{cccc}
$n_3$&$\omega=-5$&$\omega=-10$&$\omega=-20$\\
\tableline
$ F $&$-6.896  $&$ -4.530 $&$ -3.155$\\
$ 6$&$ -2.218   $&$ -2.115  $&$ -1.885 $\\
$ 15 $&$ -2.217  $&$ -2.114  $&$ -1.882 $\\
$ 21 $&$ -2.217  $&$ -2.114  $&$ -1.882 $\\
\\
$ F $&$-4.422  $&$-3.933  $&$-3.480 $\\
$6 $&$ -2.768  $&$ -2.748  $&$ -2.701 $\\
$15 $&$ -2.761  $&$ -2.744  $&$ -2.698 $\\
$21 $&$ -2.761  $&$ -2.744  $&$ -2.698 $\\
\end{tabular}
\label{tab1}
\end{table}
{\bf Figure Captions}

\vskip 0.5cm
Fig. 1 Comparison of core polarization in ordinary and halo nuclei.

Fig. 2 Diagonal matrix elements of $G_{3p1h}$ (dotted lines), $G$
(dashed lines) and $V_{\sss eff}$ (full lines) for the states
$|(p_{3/2})^2;T=1,J=0\rangle$ (upper panel) and $|(p_{1/2})^2;T=1,J=0\rangle$
(lower panel) as a function of $b_{out}$; calculations done with Paris
and Bonn-A potentials are shown by open and solid symbols, respectively.
The $G$-matrix curves are for $\omega=-5$ MeV and Pauli exclusion operator
with $(n_1,n_2,n_3)= (1,3,21)$.
\vskip 0.5cm

\begin{references}
\bibitem{ber95}  S.M. Austin and G.F. Bertsch, Scientific American
June 1995 p62.
\bibitem{al96}  J.S. Al-Khalili, Physics World June 1996 p33.
\bibitem{naz96}  W. Nazarewicz \etal,
Phys. Rev. {\bf C53} 740 (1996).
\bibitem{ham96} I. Hamamoto, H. Sagawa, and X.L. Zhang,
Phys. Rev. {\bf C53} 765 (1996);
I. Hamamoto and H. Sagawa, Phys. Rev. {\bf C53} 1492 (1996).
\bibitem{pag96} R.D. Page \etal,
Phys. Rev. {\bf C53} 660 (1996).
\bibitem{kuo96} T.T.S. Kuo, H. Muether and Azimi-Nilli, to be published
in {\em Festschrift commemorating G.E. Brown's 70th birthday}
(North-Holland Elsevier 1996).
\bibitem{lac80} M. Lacombe \etal, Phys. Rev. {\bf C21} 861 (1980).
\bibitem{mac89} R. Machleidt, Adv. Nucl. Phys. {\bf 19} (1989) 189.
\bibitem{kuo90} T.T.S. Kuo and E. Osnes, {\em Lecture Notes in Physics}
{\bf Vol.364} (Springer-Verlag 1990) p. 1.
\bibitem{kuo95} T.T.S. Kuo, E. Krmpoti\'c, K. Suzuki and R. Okamoto,
Nucl. Phys. {\bf A582}, (1995) 205.
\bibitem{shu83} J. Shurpin, T.T.S. Kuo, and D. Strottman,
Nucl. Phys. {\bf A408}, 310 (1983).
\bibitem{jorth} M. Hjorth-Jensen, T.T.S. Kuo and E. Osnes, Phys. Rep.
{\bf 261} (1995) 126.
\bibitem{tsa72} S.F. Tsai and T.T.S. Kuo, Phys. Lett. {\bf B39} 427 (1972).
\bibitem{kre76} E.M. Krenciglowa, C.L. Kung, T.T.S. Kuo and E. Osnes,
Ann. Phys. (N.Y.) {\bf 101}, 154 (1976).
\bibitem{mut92} H. M\"uther and P. Sauer, in {\em Computational Nuclear
Physics ~2}, ed. by K. Langanke, J.A. Maruhn and S. Koonin, Springer-Verlag
1992, p. 30.
\bibitem{tul95} J.K. Tuli (editor), {\em Nuclear Wallet Cards}, National
Nuclear Data Center, Brookhaven National Laboratory (1995).

\bibitem{zhu93}M.V. Zhukov \etal,
Phys. Rep. {\bf 231}, 151 (1993), and Refs. therein.
\end{references}
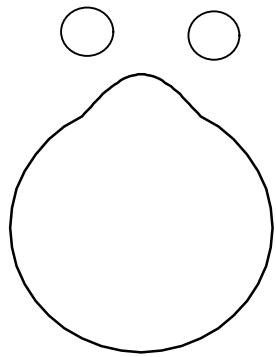
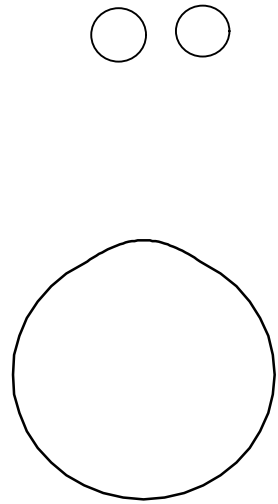
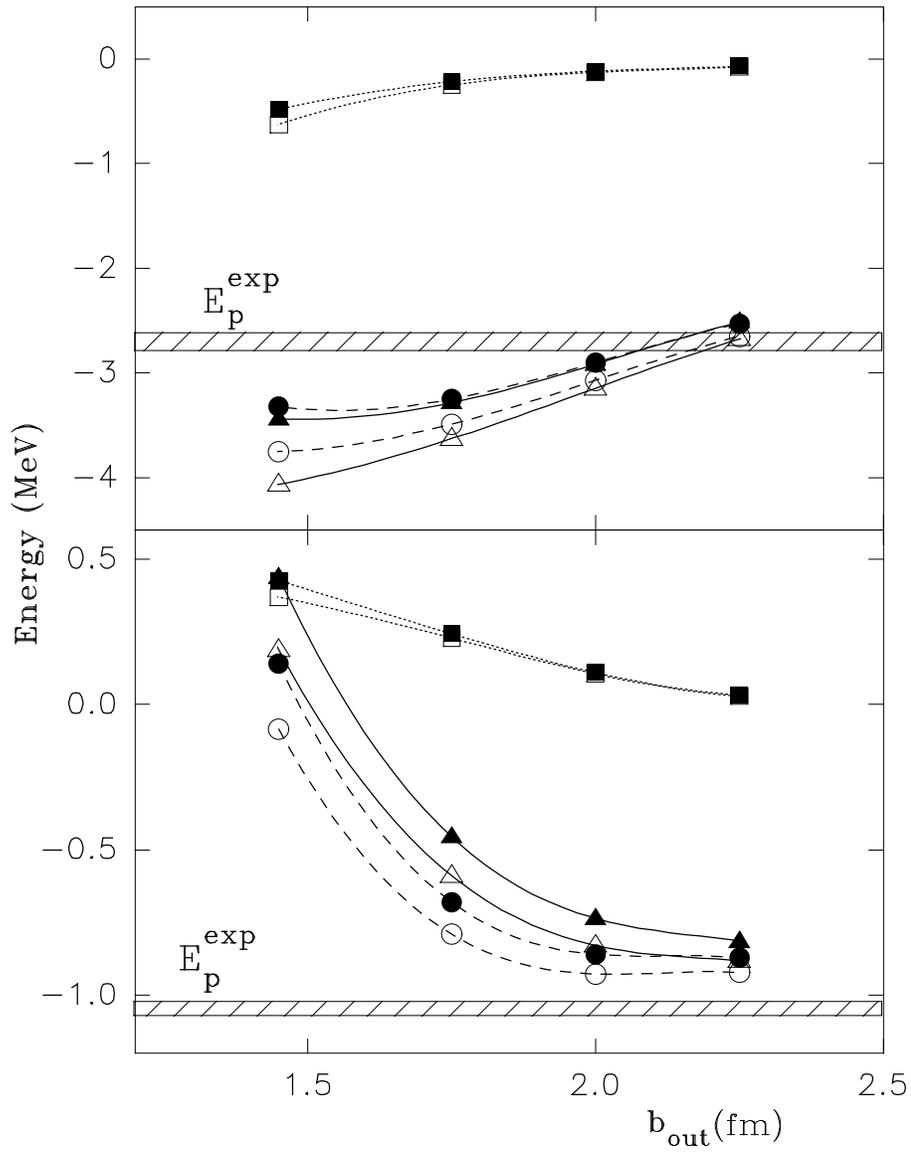
\end{document}